\documentclass[12pt]{article}
\usepackage{epsf}
\newcommand{\bP}{\mbox{\boldmath $\cal{P}$}}
\begin{document}
\begin{titlepage}
\title{Penguin amplitudes in $B^{+} \to \pi^+ K^{*0},  K^+ \bar{K}^{*0}$ decays}
\author{
{Piotr \.Zenczykowski }\\
{\em Division of Theoretical Physics},\\
{\em the Henryk Niewodnicza\'nski Institute of Nuclear Physics,}\\
{\em Polish Academy of Sciences,}\\
{\em Radzikowskiego 152,
31-342 Krak\'ow, Poland}\\
}
\maketitle
\begin{abstract}
The question of the relative size of two independent penguin amplitudes 
is studied using the data on the
$B^{+} \to \pi^{+} K^{*0} $, 
and $B^+ \to K^+\bar{K}^{*0}$ decays. 
Our discussion involves a Regge-phenomenology-based estimate of $SU(3)$
breaking in the final quark-pair-creating hadronization process.
The results are in agreement with earlier estimates of the relative size of the
two penguins obtained from $B^+ \to \pi^+ K^0, K^+ \bar{K}^0$.

\end{abstract}
\vfill
\end{titlepage}

\section{Introduction}
Rare charmless nonleptonic B meson decays give us  a lot of information
 on the weak interactions of quarks. 
 Yet, since quarks are forever confined, 
 this information is necessarily blended with strong interaction effects. 
A proper and reliable understanding of the latter is crucial for the extraction
of New Physics effects (if any).
At present, however, the only way to achieve such an understanding 
 is through the
parametrization of strong interaction effects and the
subsequent extraction of relevant parameters from the experimental data.

For example, it is known that the effective tree ($\tilde{T}$) 
and colour-suppressed ($\tilde{C}$)
amplitudes involve contributions from some of the penguin amplitudes as well
(see eg. \cite{GHLR,Buras2004,Chiang2004}).
Disentangling this penguin contribution from the `true' $C$ and $T$ amplitudes
would constitute part of the proper understanding of `standard physics', both
at the quark- and the hadron-level, and a check on theoretical predictions at
the quark
level.
Unfortunately, achieving this goal
is not possible on the basis of $B\to \pi\pi$ data alone, which permit an
extraction of $\tilde{C}/\tilde{T}$, but not $C/T$. 
Indeed, estimating the latter
would require knowledge of
the size of one of the two independent penguin amplitudes \cite{Buras2004}.
Knowing the branching ratio of a single purely-penguin process is not sufficient here.
If the two penguin contributions
are to be separated properly,
additional data and some related assumptions are needed.

The best place to analyse the size of penguin contributions should be
 in those processes 
 in which only penguin amplitudes are present.
For this reason, ref. \cite{ZenAPP2010} considered $B^+ \to \pi^+ K^0$ decays in
conjunction with $B^+ \to K^+ \bar{K}^0$. In those two processes, 
amplitudes $C$ and $T$ are absent and only 
penguin
amplitudes contribute. As shown in \cite{ZenAPP2010}, 
under reasonable assumptions concerning SU(3) symmetry breaking,
it is then possible to extract from the experimental data 
the relative size and phase of the two independent
penguin amplitudes, and therefore estimate
the effect of the penguin-induced corrections. 

The present paper is concerned with the extraction of
the ratio of two similar penguin amplitudes from the related, but independent,
decays $B^+ \to \pi^+ K^{*0}$ and $B^+ \to K^+ \bar{K}^{*0}$.
Our results are fully consistent with the findings of ref. \cite{ZenAPP2010}.

\section{Penguin amplitudes}
For future comparison, we first briefly recall the notation used in
\cite{ZenAPP2010} for the description of penguins in $B \to PP$
decays, and then introduce an analogous notation for $B \to PV$ processes.

\subsection{$B \to PP$}
Ref. \cite{ZenAPP2010} was concerned with the extraction of independent penguin
amplitudes relevant for $B \to PP$ decays. In that case,
when the elements of the CKM matrix were factored out, the
three independent strong penguin amplitudes corresponding to internal $k$-quark loops ($k=u,c,t$) 
were denoted in \cite{ZenAPP2010}, \cite{SZ2005} by ${\cal{P}}_k$.  
Because of the unitarity of the CKM matrix only
two combinations of these amplitudes can be extracted
 from experimental data.
For $b \to
d$ transitions we choose the following two combinations ($q=u,c$)
\footnote{In order to make our notation
 more transparent, 
all amplitudes denoted in \cite{ZenAPP2010} and \cite{SZ2005} by calligraphic
letters are written here in bold.}: 
\begin{equation}
\label{Pq}
P_q\equiv - \lambda^{(d)}_c {\bP}_{tq}=A\lambda^3{\bP}_{tq},
\end{equation}
where
\begin{equation}
\label{PenguindiffPP}
{\bP}_{tq}={\bP}_t-{\bP}_q,
\end{equation}
is the difference of two independent penguin amplitudes,
and $\lambda^{(k)}_q$ is given in terms of the CKM matrix $V$:
\begin{equation}
\lambda^{(k)}_q=V_{qk}V^*_{qb},
\end{equation}
with $A$ and $\lambda=0.225$ being the Wolfenstein parameters
(we do not need the actual value of $A$ in our calculations).

If one assumes that the difference between the $b\to d$ and $b \to s$
penguin amplitudes arises solely from different CKM factors,
the $b \to s$ transitions may be
expressed in terms of the same strong penguin amplitudes ${\bP}_{tq}$
as the $b \to d$ processes.
This assumption ignores any possible dependence on the spectator quark and on the
flavour of the additional quark-antiquark pair which 
appears in the final state and has to be produced via strong
 interactions. 
 While the spectator-independence of the amplitude seems a fairly reasonable assumption,
  the production of the additional quark-antiquark pair may be strongly
 flavour-dependent, as discussed at length in \cite{ZenAPP2010}.
 We will come to this point later. For the $b \to s$ processes, 
the corresponding products of strong penguin amplitudes and CKM
factors will be denoted with primed letters, i.e.:
\begin{equation}
\label{Pprimq}
P'_q\equiv - \lambda^{(s)}_c {\bP}_{tq},
\end{equation}
so as to distinguish them from (unprimed) amplitudes $P_q$ used for $b \to d$ decays.

We are interested in the relative size of the two independent combinations of 
penguin amplitudes, i.e. in the ratio:
\begin{equation}
{\bP}_{tu}/{\bP}_{tc}.
\end{equation}
In order to discuss this ratio it
is convenient to introduce \cite{ZenAPP2010}:
\begin{equation}
\label{znadzeta}
ze^{i\zeta}\equiv R_b \frac{P_u}{P_c}=R_b \frac{P'_u}{P'_c}=R_b 
\frac{{\bP}_{tu}}{{\bP}_{tc}},
\end{equation}
where \cite{UTFit}
\begin{equation}
R_b=\sqrt{\bar{\rho}^2+\bar{\eta}^2}=-\frac{\lambda^{(d)}_u}{\lambda^{(d)}_c}
e^{-i\gamma}\approx 0.38.
\end{equation}
In our calculations we use the central value of the
SM prediction $\gamma = 69.6^o\pm 3.1^o$, which compares very well
with the value of $69.8^o \pm 3.0^o$ fitted in \cite{UTFit}.

\subsection{$B \to PV$}
For the $B \to PV $ decays there are two possible cases: the spectator
quark (i.e. the quark not involved in weak interactions) ends up either in a pseudoscalar
($P$) or in a vector ($V$) meson. In order to distinguish between 
these two cases, 
we introduce the subscript $P$ or $V$ in all relevant amplitudes
(hence the corresponding amplitudes
 are distinguished 
from the $B \to PP$ amplitudes for which we do not use a subscript) \cite{SZ2005}.
For example, $P_P$ ($P_V$) denotes full penguin amplitudes relevant in $B \to PV$
decays with 
spectator quark ending in a $P$ ($V$) meson.

For the decays with the spectator quark emerging in the final pseudoscalar
meson, and with the CKM matrix elements factored out, we have again three independent strong penguin amplitudes 
${\bP}_{P,k}$ corresponding to $k$-quark running in an internal loop.
As in the case of $B \to PP$ decays, we use the unitarity of the CKM matrix to
describe the relevant total penguin amplitudes in terms of two combinations
($q=u,c$) analogous to $\bP_{tq}$ of Eq. (\ref{PenguindiffPP}):
\begin{equation} 
{\bP}_{P,tq}={\bP}_{P,t}-{\bP}_{P,q}.
\end{equation} 
The full penguin amplitudes with spectator quark ending in a pseudoscalar meson
are then
\begin{eqnarray}
\label{PP}
P_P&=&-\lambda_u^{(d)}{\bP}_{P,tu}-\lambda_c^{(d)}{\bP}_{P,tc},\\
\label{PprimP}
P'_P&=&-\lambda_u^{(s)}{\bP}_{P,tu}-\lambda_c^{(s)}{\bP}_{P,tc},
\end{eqnarray}
for $b \to d$ and $b\to s$ transitions respectively.
Again, the above formulae ignore any other possible dependence on the spectator quark
and any dependence on the flavour of the quark-antiquark pair that has to be additionally
produced via strong interactions. 

In order to discuss the ratio of penguin amplitudes
${\bP}_{P,tu}/{\bP}_{P,tc}$ relevant in $B \to PV$ decays, and to compare it
later with the $B \to PP$ case, it is
convenient to define the counterpart of Eq. (\ref{znadzeta}):
\begin{equation}
z_Pe^{i\zeta_P}\equiv R_b \frac{P_{P,u}}{P_{P,c}}=R_b \frac{P'_{P,u}}{P'_{P,c}}
=R_b 
\frac{{\bP}_{P,tu}}{{\bP}_{P,tc}},
\end{equation}
where $P_{P,q}$ and $P'_{P,q}$ are defined analogously to
Eqs (\ref{Pq},\ref{Pprimq}):
\begin{eqnarray}
\label{PPq}
P_{P,q}&\equiv& - \lambda^{(d)}_c {\bP}_{P,tq},\\
P'_{P,q}&\equiv& - \lambda^{(s)}_c {\bP}_{P,tq}.
\end{eqnarray}

\section{Decays $B^+ \to \pi^+ K^{*0}, K^+\bar{K}^{*0}$}
If there is no dependence on the flavour of the additional quark-antiquark pair
produced via strong final state interactions,
the amplitudes for the $B^+ \to \pi^+ K^{*0}$ and $B^+ \to K^+\bar{K}^{*0}$ decays are:
\begin{eqnarray}
A(B^+\to\pi^+K^{*0})&=&P'_P,\\
A(B^+\to K^+\bar{K}^{*0})&=&P_P.
\end{eqnarray}
Using Eqs (\ref{PP},\ref{PprimP}) and the relation
\begin{equation}
P'_{P,c}=\frac{\lambda^{(s)}_c}{\lambda^{(d)}_c} P_{P,c}=-\frac{1}{\sqrt{\epsilon}}P_{P,c},
\end{equation}
where
\begin{equation}
\epsilon=\frac{\lambda^2}{1-\lambda^2}\approx 0.053,
\end{equation}
one can
reexpress the above amplitudes in terms of $P_{P,c}$, $z_P$ and $\zeta_P$
as:
\begin{eqnarray}
\label{piK*}
A(B^+\to\pi^+K^{*0})&=&-\frac{1}{\sqrt{\epsilon}}P_{P,c}
(1+\epsilon z_P e^{i(\zeta_P+\gamma)}),\\
\label{KK*SU3}
A(B^+\to K^+\bar{K}^{*0})&=&P_{P,c}(1-z_P e^{i(\zeta_P+\gamma)}).
\end{eqnarray}
The above formulae are
completely analogous to those relevant for
$B^+ \to \pi^+ K^0$ and $B^+ \to K^+ \bar{K}^0$ decays \cite{ZenAPP2010}
(with the replacements $P_{P,c}\to P_c$, $z_P\to z$, and
$\zeta_P \to \zeta$). 
 
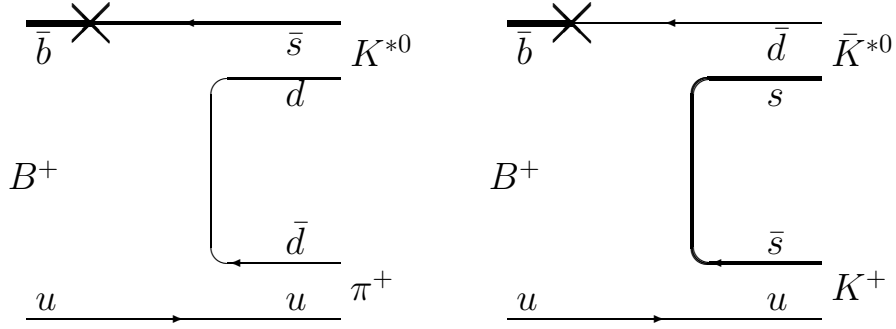
\begin{figure}
\caption{$B^+$ decays to $\pi^+ K^{*0}$ and $K^+\bar{K}^{*0}$ final states. Penguin $\bar{b} \to
\bar{s}$ and $\bar{b} \to \bar{d}$ transitions are
denoted with crosses. 
}
\label{fig1}
\begin{center}
\setlength{\unitlength}{0.7pt}
  \begin{picture}(460,220)
  \put(0,-20){\begin{picture}(200,220)
  \put(10,50){\vector(1,0){85}}
  \put(180,50){\line(-1,0){85}}
  \put(180,80){\vector(-1,0){60}}
  \put(110,90){\line(0,1){80}}
  
  \put(10,210){\line(1,0){85}}
  \put(10,209.5){\line(1,0){85}}
  \put(10,210.5){\line(1,0){85}}
  \put(10,209){\line(1,0){35}}
  \put(10,211){\line(1,0){35}}
  \put(10,208.5){\line(1,0){35}}
  
  \put(35,200){\line(1,1){20}}
  \put(35,200.5){\line(1,1){20}}
  \put(35,199.5){\line(1,1){20}}
  \put(35,201){\line(1,1){20}}
  
  \put(35,220){\line(1,-1){20}}
  \put(35,220.5){\line(1,-1){20}}
  \put(35,219.5){\line(1,-1){20}}
  \put(35,219){\line(1,-1){20}}
  
  \put(180,210){\vector(-1,0){85}}
  \put(180,210.5){\line(-1,0){85}}
  \put(180,209.5){\line(-1,0){85}}
  \put(120,180){\line(1,0){60}}
  \put(120,90){\oval(20,20)[bl]}
  \put(120,170){\oval(20,20)[tl]}
  \put(185,187){\makebox{\large {${K^{*0}}$}}}
  \put(185,60){\makebox{\large {${\pi^+}$}}}
  \put(150,193){\makebox{\large {$\bar{s}$}}}
  \put(15,188){\makebox{\large {$\bar{b}$}}}
  \put(150,165){\makebox{\large {${d}$}}}
  \put(150,85){\makebox{\large {$\bar{d}$}}}
  \put(150,55){\makebox{\large {${u}$}}}
  \put(15,55){\makebox{\large {${u}$}}}
  \put(0,120){\makebox{\large {${B^+}$}}}
  \end{picture}}
  
  \put(260,-20){\begin{picture}(200,220)
  \put(10,50){\vector(1,0){85}}
  \put(180,50){\line(-1,0){85}}
  \put(180,80){\vector(-1,0){60}}
  \put(180,80.5){\line(-1,0){60}}
  \put(180,79.5){\line(-1,0){60}}
  
  \put(10,208.5){\line(1,0){35}}
  \put(10,209){\line(1,0){35}}
  \put(10,210){\line(1,0){35}}
  \put(10,211){\line(1,0){35}}
  \put(10,209.5){\line(1,0){35}}
  \put(10,210.5){\line(1,0){35}}
  
  \put(35,200){\line(1,1){20}}
  \put(35,200.5){\line(1,1){20}}
  \put(35,199.5){\line(1,1){20}}
  \put(35,201){\line(1,1){20}}
  
  \put(35,220){\line(1,-1){20}}
  \put(35,220.5){\line(1,-1){20}}
  \put(35,219.5){\line(1,-1){20}}
  \put(35,219){\line(1,-1){20}}

  \put(110,90){\line(0,1){80}}
  \put(110.5,90){\line(0,1){80}}
  \put(109.5,90){\line(0,1){80}}
  \put(10,210){\line(1,0){85}}
  \put(180,210){\vector(-1,0){85}}
  \put(120,180){\line(1,0){60}}
  \put(120,179.5){\line(1,0){60}}
  \put(120,180.5){\line(1,0){60}}
  \put(120,90){\oval(20.5,20.5)[bl]}
  \put(120,90){\oval(19,19)[bl]}
 \put(120,90){\oval(20,20)[bl]}
  \put(120,90){\oval(21,21)[bl]}
  \put(120,170){\oval(20,20)[tl]}
  \put(120,170){\oval(19,19)[tl]}
  \put(120,170){\oval(20.5,20.5)[tl]}
  \put(120,170){\oval(21,21)[tl]}
  \put(185,187){\makebox{\large {$\bar{K}^{*0}$}}}
  \put(185,60){\makebox{\large {${K^+}$}}}
  \put(150,165){\makebox{\large {${s}$}}}
  \put(150,85){\makebox{\large {$\bar{s}$}}}
  \put(15,188){\makebox{\large {$\bar{b}$}}}
  \put(150,190){\makebox{\large {$\bar{d}$}}}
  \put(150,55){\makebox{\large {${u}$}}}
  \put(15,55){\makebox{\large {${u}$}}}
  \put(0,120){\makebox{\large {${B^+}$}}}
  \end{picture}}
\end{picture}
\end{center}
\end{figure}

In reality, the decays $B^+ \to \pi^+ K^{*0}$ and $B^+ \to K^+ \bar{K}^{*0}$
may differ in an important way, not taken into account in the above formulae.
Indeed, the newly created quark-antiquark pair is produced 
(long after the weak decay)
by long-range strong interactions (see Fig. \ref{fig1}). Now, it is known that in such
strong-interaction processes - in which 
newly produced quarks $q$ and $\bar{q}$ end up in different hadrons - 
the production of the $s\bar{s}$ pair is strongly
suppressed at high energies when compared to that of the light $q\bar{q}$
pair.
In order to take this effect into account, we modify
formula (\ref{KK*SU3}) by introducing a corresponding
suppression factor $\kappa$:
\begin{equation}
\label{KK*}
A(B^+\to K^+\bar{K}^{*0})=\kappa P_{P,c}(1-z_P e^{i(\zeta_P+\gamma)}).
\end{equation}
A reliable estimate of this suppression factor may be obtained using Regge
phenomenology as discussed in much detail in ref. \cite{ZenAPP2010}, where it was
shown that
\begin{equation}
\kappa=(m^2_B/s_0)^{\alpha_{0}(K^*)-\alpha_0(\rho)},
\end{equation}
with $s_0 \approx 1~GeV^{-2}$, and $\alpha_{0}(M)$ being the intercept of Regge
trajectory corresponding to meson $M$. 
Since the values of the intercepts are extracted directly from high
energy scattering experiments, 
their values take into account  
{\it all} final state interactions.
For $m_B^2 =27.9 ~GeV^2$ one finds then that
$\kappa \approx 0.5-0.6$ \cite{ZenAPP2010,Irving,Martin}. In the following we take $\kappa =
0.55.$

\subsection{Extraction of penguins' relative sizes and phases}
The CP-averaged branching ratios for the 
$B^+ \to \pi^+K^{*0}, K^+\bar{K}^{*0}$ decays are
given by
\begin{eqnarray}
\langle {\cal{B}}(B^+ \to \pi^+ {K}^{*0}) \rangle_{CP}&\approx &
\frac{1}{\epsilon}|P_{P,c}|^2\,(1+2\,\epsilon\, z_P\, \cos \zeta_P \cos \gamma),\nonumber\\
\label{BRatios}
\langle {\cal{B}}(B^+ \to K^+ \bar{K}^{*0}) \rangle_{CP}&=& \kappa ^2
|P_{P,c}|^2\,(1+z_P^2-2z_P \cos \zeta_P \cos \gamma),
\end{eqnarray} 
where we neglected the terms of order $(\epsilon z_P)^2$. 
The latter assumption is well justified since
$\epsilon \approx 1/20$ and 
\begin{equation}
z_P=0.38\, \left|\frac{\bP_{P,t}-\bP_{P,u}}
{\bP_{P,t}-\bP_{P,c}}\right|
\end{equation}
is expected to be of order $1$ (or smaller if the top quark penguin
$\bP_{P,t}$ dominates).

We now take the ratio of the two branching fractions to 
find that
\begin{eqnarray}
\label{conditionRKKpiK}
R^{KK^*}_{\pi K^*}\equiv \frac{\langle {\cal{B}}(B^+ \to K^+ \bar{K}^{*0}) \rangle_{CP}}
{\langle {\cal{B}}(B^+ \to \pi^+ {K}^{*0}) \rangle_{CP}}
&=&\epsilon \,\kappa ^2\,
\frac{1+z_P^2-2z_P \cos \zeta_P \cos \gamma}{1+2\,\epsilon\, z_P\, 
\cos \zeta_P \cos \gamma}
\end{eqnarray}
(thus, we do not need the value of $|P_{P,c}|$).

The averages of experimental branching ratios for $B^+ \to \pi^+ K^{*0}$ and
$B^+ \to K^+\bar{K}^{*0}$ decays are (in units of $10^{-6}$) \cite{HFAG}:
\begin{eqnarray}
\langle {\cal{B}}(B^+ \to \pi^+ {K}^{*0}) 
\rangle_{CP}&=&9.9 ^{+0.8}_{-0.9},\nonumber\\
\label{BRexp}
\langle {\cal{B}}(B^+ \to K^+ \bar{K}^{*0})
\rangle_{CP}&=&0.68\pm 0.19.
\end{eqnarray}
corresponding to $R^{KK^*}_{\pi K^*}=0.069 \pm 0.020$.

For the asymmetries $A_{CP}$ we find:
\begin{eqnarray}
\label{asympiK*}
{A}_{CP}(B^+\to\pi^+K^{*0})&=&\frac{2\, \epsilon z_P \,\sin \zeta_P \sin \gamma}
{1+2\, \epsilon z_P \cos \zeta_P \cos \gamma},\\
{A}_{CP}(B^+\to K^+\bar{K}^{*0})&=&-\,\,\frac{2\,z_P\sin\zeta_P\sin\gamma}
{1+z_P^2-2\,z_P\cos\zeta_P\cos\gamma},
\end{eqnarray}
while the experimental value is known for the first process only, with the
average \cite{HFAG}:
\begin{equation}
\label{CPpiK*exp}
A_{CP}(B^+\to\pi^+K^{*0})=-0.038\pm 0.042.
\end{equation}

\begin{figure}[t]
\caption{Contour plot showing the experimentally allowed
region of ($z_P$, $\zeta_P$) plane. 
Approximately vertical (horizontal) lines correspond to 
branching ratio (asymmetry) constraints.
Solid lines represent constraints for central
experimental values, dashed lines - for one standard deviation.
}
\label{fig2}
\begin{center}
\mbox{\epsfbox{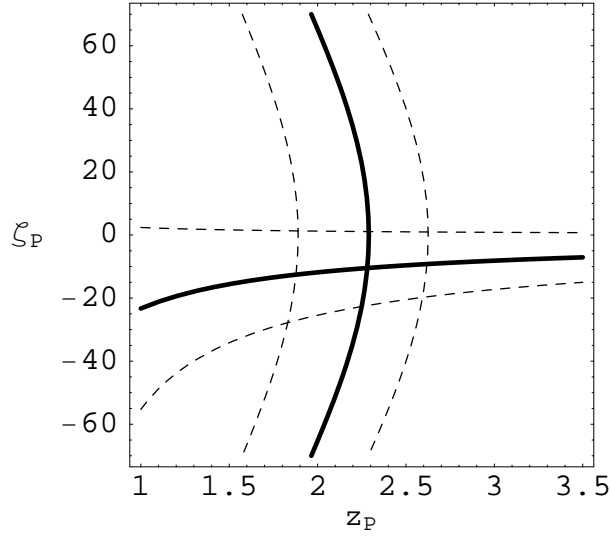}}
\end{center}
 \end{figure}

Equations (\ref{conditionRKKpiK},\ref{BRexp},\ref{asympiK*},\ref{CPpiK*exp})
provide two conditions on $z_P$ and $\zeta_P$.
Their solution yields:
\begin{eqnarray}
z_P&=&2.3\pm 0.3,\nonumber\\
\label{solution1}
\zeta_P&=&-11^o \pm 12^o,
\end{eqnarray}
or
\begin{eqnarray}
z_P&=&1.45^{+0.30}_{-0.40},\nonumber\\
\label{solution2}
\zeta_P&=&-165^o \pm 15^o.
\end{eqnarray}
The second solution corresponds to
${\bP}_{P,tu}$ and ${\bP}_{P,tc}$ with nearly opposite phases.
This is highly unlikely since the masses of $u$ and $c$ quarks are small when
compared to the mass of $t$ quark, and - consequently - the two penguins
should have similar strong phases.
Therefore, this solution may be discarded.
For the first solution, the relevant contour plot in the ($z_P$, $\zeta_P$) 
plane is shown in Fig. \ref{fig2}.

The values of $z_P$, $\zeta_P$  
 extracted from experimental data on $B \to \pi K^*, KK^*$ decays
 (Eq.(\ref{solution1}))
are very similar to the values
of
\begin{eqnarray}
z&=&1.8-2.3,\nonumber\\
\zeta&=&-15^o ~{\rm to}~0^o,
\end{eqnarray}
obtained in \cite{ZenAPP2010} in
a similar analysis of the
 $B \to \pi K, KK$ decays.
 
 This agreement corroborates the idea that there is no essential
 difference between the short-range penguin transitions 
 in $B \to PP$ and $B \to PV$ decays.
 The only important difference between the pure penguin-driven processes 
 in $B \to PP$ and $B \to PV$ processes arises from final-state long-range
 interactions. 
 In particular, the relative size of the relevant branching ratios
 indicates that one has to include the effect of $SU(3)$ breaking ($\kappa \ne 1$) in the
 final $q\bar{q}$-creation process.

\section{Discussion and conclusions}
It is  straightforward to derive
 the following formula for the $A_{CP}(B^+\to K^+\bar{K}^{*0})$ asymmetry:
 \begin{equation}
 \label{AsymKplusK0star}
 A_{CP}(B^+\to K^+\bar{K}^{*0})=
 -\kappa^2 \frac{\langle {\cal{B}}(B^+ \to \pi^+ {K}^{*0}) \rangle_{CP}}
 {\langle {\cal{B}}(B^+ \to K^+ \bar{K}^{*0})
\rangle_{CP}} A_{CP}(B^+\to\pi^+K^{*0}).
 \end{equation}

For $\kappa=0.55$ one has
\begin{equation}
A_{CP}(B^+\to K^+\bar{K}^{*0})\approx +0.17 \pm 0.19,
\end{equation}
with the dominant contribution to the error coming from 
$A_{CP}(B^+\to \pi^+K^{*0})$.
The asymmetry $A_{CP}(B^+\to K^+\bar{K}^{*0})$ is expected to be some four times
larger than $A_{CP}(B^+\to \pi^+K^{*0})$.
This is because it is not proportional to the suppression factor of $\epsilon$:
the two contributing penguin amplitudes are of comparable sizes.
Consequently, it would be interesting to have $A_{CP}(B^+\to K^+\bar{K}^{*0})$ measured 
as it is particularly sensitive
to the size of the
relative strong phase of the two penguin contributions.\\

We have found that the ratios of the two penguin amplitudes 
are similar in the
$B^+ \to \pi^+ K^0 ~(K^+\bar{K}^0)$ and $B^+ \to \pi^+ K^{*0}~ (K^+\bar{K}^{*0})$
sectors, i.e. that they are independent
of the final state, 
and that the SU(3) breaking factor $\kappa$ is universal. Consequently,
it is natural to expect that the same will be true for the
$B^+ \to  \rho^+ K^0$ ($K^{*+} \bar{K}^0$) sector, where
the spectator quark ends up in a vector meson
(i.e. that $z_P \approx z \approx z_V$, 
and $\zeta_P \approx \zeta \approx \zeta_V$).

This leads us to expect that the branching fraction of the
$B \to K^{*+}\bar{K}^0$ decay should be approximately equal to
\begin{equation}
\langle{\cal{B}}(B^+ \to K^{*+}\bar{K}^0)\rangle_{CP}=
\langle{\cal{B}}(B\to \rho^+ K^0)\rangle_{CP}
\frac{\langle{\cal{B}}(B^+\to K^+\bar{K}^{*0})\rangle_{CP}}
{\langle{\cal{B}}(B^+\to \pi^+ K^{*0})\rangle_{CP}} \approx
0.55 \pm 0.2.
\end{equation}
An analogon of Eq. (\ref{AsymKplusK0star}) may also be written.

It is hoped that future experimental work on $B^+ \to K^{*+}\bar{K}^0$,
$B^+\to K^+\bar{K}^{*0}$, $B^+\to\pi^+K^{*0}$,
and $B\to \rho^+ K^0$ processes will further 
corroborate the ideas of this paper.

\section{Acknowledgements}
I would like to thank Leonard Le\'sniak for reading the manuscript prior to
publication.
This work has been partially supported by the Polish Ministry of Science and
Higher Education research project No N N202 248135.

\vfill

\vfill

\end{document}